\documentclass{astlb}

\usepackage[T2A]{fontenc}
\usepackage[utf8]{inputenc}
\usepackage[russian,english]{babel}

\usepackage{lscape}
\usepackage{booktabs} 
\usepackage{multicol,lipsum}
\usepackage{xspace} 

\usepackage{ragged2e}
\usepackage{color}

\usepackage{graphicx}	% Including figure files
\usepackage{amssymb}
\usepackage{amsmath}	% Advanced math s commands
\usepackage{longtable}

\newcommand{\fu}{erg~s$^{-1}$~cm$^{-2}$}

\newcommand{\xmm}{XMM-{Newton}}

\def\ps1{{Pan-STARRS1}}

\def\20{{L20}}

\def\deg{\hbox{$^\circ$}}
\def\arcmin{\hbox{$^\prime$}}
\def\arcsec{\hbox{$^{\prime\prime}$}}

\begin{document}

\journalinfo{(2023)}{49}{11}{655}[665]
%\UDK{524.77}

\title{SRG/ART-XC Galactic Plane Survey near Galactic Longitude $l\simeq20^\circ$: Catalog of Sources}

\author{
D. I.~Karasev\address{1}\email{dkarasev@cosmos.ru},
A.~N.~Semena\address{1},
I.~A.~Mereminskiy\address{1}
A.~A.~Lutovinov\address{1},
R.~A.~Burenin\address{1},
R.~A.~Krivonos\address{1},
S.~Yu.~Sazonov\address{1},
V.~A.~Arefiev\address{1},
M.~V.~Buntov\address{1},
I.~Yu.~Lapshov\address{1}
V.~V.~Levin\address{1},
M.~N.~Pavlinsky\address{1},
A.~Yu.~Tkachenko\address{1},
A.~E.~Shtykovsky\address{1}
\addresstext{1}{\it Space Research Institute, Russian Academy of Sciences, Moscow, 117997 Russia}
}

\shortauthor{Karasev et al.}

\shorttitle{SRG/ART-XC L20 field survey}

\submitted{20.10.2023 \\
Revised October 20, 2023; Accepted November 21, 2023}

\begin{abstract}
We present a catalog of sources detected by the Mikhail Pavlinsky ART-XC telescope onboard the SRG space observatory during the observations of the Galactic plane region near a longitude  $l\simeq20$\deg\ (L20 field) in October 2019. The L20 field was observed four times in the scanning mode, which provided a uniform coverage of the sky region with a total area of $\simeq24$ sq. deg with a median sensitivity of 
$8\times10^{-13}$ erg s$^{-1}$ cm$^{-2}$ (at 50\% detection completeness) in the 4--12 keV. As a result, we have detected 29 X-ray sources at a statistically significant level, 11 of which have not been detected previously by other X-ray observatories. Preliminary estimates show that four of them can presumably be extragalactic in nature. We also show that the source SRGA\,J183220.1$-$103508 (CXOGSG\,J183220.8$-$103510), is most likely a galaxy cluster containing a bright radio galaxy at redshift $z\simeq0.121$.  


\end{abstract}

\section*{INTRODUCTION}

One of the most interesting sky regions for a study is the Galactic plane populated by a large number of X-ray sources of various nature. Among them are cataclysmic variables, X-ray binaries (low-mass and high-mass ones), and extended X-ray sources, such as supernova remnants, pulsar nebulae, etc. Deep observations of the Galactic plane allow one to detect more and more X-ray sources in our Galaxy \citep[see, e.g., the INTEGRAL surveys,][]{2004AstL...30..534M,2006AstL...32..145R,2006ApJ...636..765B,2012A&A...545A..27K,2017MNRAS.470..512K,Semena_bulge} and to study the statistical and spatial distributions of objects of various classes. 

In 2019, after the launch of the Spectrum--Roentgen--Gamma (SRG, \citealt{srg}) observatory, before proceeding to the accomplishment of its main task, an all-sky survey, it was necessary to test and study the characteristics of the SRG instruments in real conditions. During this calibration and performance verification (CalPV) phase, in particular, it was required to ascertain the capabilities of the telescopes to study faint X-ray sources located near the Galactic plane. The test sky field to accomplish this task was selected in such a way that it was far from too bright X-ray sources (that could degrade significantly the background conditions when detecting faint sources) and regions with significant interstellar absorption, which is critically important for observations with the eROSITA telescope \citep{2021A&A...647A...1P} operating in the soft X-ray energy band.   
As a result, we chose a $\simeq$6\deg$\times$4\deg\ Galactic field with center coordinates $l\simeq20^{\circ}, b\simeq0^{\circ}$ that we will call below the L20 field.
%(RA=18:28:06.57 Dec=-11:16:04.5) 

Although the CalPV program is primarily designed to verify the performance of the instruments, these data are also suitable for scientific studies. In this paper we present the results of a deep survey of the L20 Galactic plane region based on the data from the Mikhail Pavlinsky ART-XC telescope  \citep{2021A&A...650A..42P}, obtained in October 2019.

\section{OBSERVATIONS}
\label{sec:obs}

%================================================
\begin{table}
    \caption{List of scanning observations of the L20 field with the ART-XC telescope during the CalPV phase}
    \label{tab:l20_scans}
    \centering
    \vskip 2mm
    \renewcommand{\arraystretch}{1.5}
    \renewcommand{\tabcolsep}{0.15cm}
    \scriptsize
    \begin{tabular}{llcccc}
\toprule
    \hline
{} & Obs. ID  & TSTART & RA  & Dec  & Exp. \\
{} &         & UTC    & deg & deg  & ks  \\
\midrule
    \hline
1 & 50000600100 & 2019-10-13 05:59 & 274.6405 & -13.1485 &        50  \\
2 & 50000700100 & 2019-10-13 20:48 & 274.6384 & -13.0986 &        50  \\
3 & 50000800100 & 2019-10-15 16:52 & 274.6362 &  -13.0486 &        50 \\
4 & 50000900100 & 2019-10-16 07:42 & 274.6340 & -12.9987 &        50  \\
\bottomrule
    \hline

\end{tabular}

\end{table}
%================================================

The ART-XC grazing-incidence X-ray telescope consists of seven identical modules sensitive in the 4--30 keV energy band (with the maximum of the effective area in the 7--11 keV energy band). The field of view is 36\arcmin\ in diameter, which allows wide-field sky surveys to be conducted at relatively high energies with a high sensitivity and homogeneity. The high stability of the instrumental background due to the location of the SRG observatory at the Lagrange point L2 of the Sun–Earth system and the detector protection against the stray light from X-ray photons are additional advantages of the telescope \citep{2021A&A...650A..42P}.

During the calibration and performance verification phase the ART-XC telescope observed the L20 field for about 200 ks in the scanning mode. Owing to this mode, the SRG observatory can carry out the studies of sky fields whose sizes exceed considerably the sizes of the field of view of its telescopes. This is done with a slow “serpentine” passage through the required region in the sky \citep{2021A&A...650A..42P}. In addition, this mode of observations is favorable from the standpoint of exposure uniformity and background quality.

\begin{figure*}
    \centering
    \includegraphics[width=1.2\columnwidth,trim={1.2cm 6.7cm 1.5cm 6.5cm},clip]{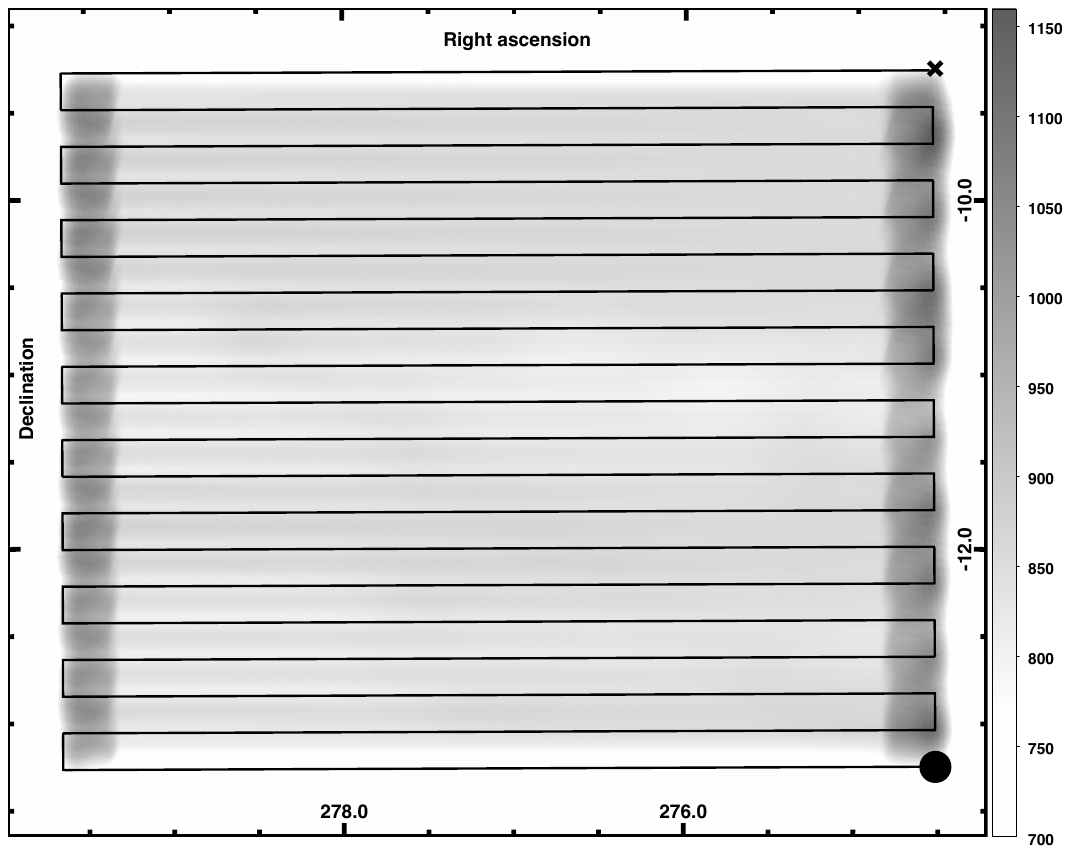}
    \caption{Exposure map of the L20 Galactic field survey performed with the ART-XC telescope. The black line indicates the trajectory (“serpentine”) of the center of the ART-XC field of view during the scanning observations. The black circle and the cross mark the scanning start and end points, respectively.}
    \label{fig:expo}
\end{figure*}

During the observations of the L20 field four scans were made in the period from October 13 to 16, 2019 (see Table \ref{tab:l20_scans} and Fig. \ref{fig:expo}). A similar $\simeq$6\deg$\times$4\deg\, sky region was covered in each scan, while the trajectories of the ART-XC field of view in different observations were repeated, except for the small  $\simeq$1.5\arcmin\ shift in declination. Owing to the chosen strategy, the L20 field $\simeq24$ sq. deg in area was covered with a median effective exposure time of $\simeq830$ s.  

%================================================
%================================================

\section{DATA PROCESSING}
\label{sec:survey}

For the detection of sources in the field under study, we used a method based on a comparison of the probabilities of the observed ensemble of events for the model including the background and a point source at a specified position in the sky or only the background that is optimal for the choice of a model \citep[][]{Semena_algo}. This algorithm has been successfully applied previously for the detection of sources in the Galactic bulge field \citep{Semena_bulge} and is also consistent with the processing methods applied in producing the catalog of sources of the ART-XC all-sky survey \citep{2022A&A...661A..38P}.

When selecting the sources in the field under study, we were guided primarily by the achievement of a high purity of the catalog being produced, i.e., a low fraction of false sources in it. For this purpose, we performed Monte Carlo simulations of an empty field, with a particle background, a sky background, and an exposure time equivalent to the L20 field survey. As a result, we chose the constant detection threshold (in terms of the likelihood function) $\Delta \ln{L} > 11.4$, that provides the expected number of false sources $N_{\rm false}=1$ and a constant density of false sources in the entire field (for more details, see \citet{Semena_algo}). 

\begin{figure}
    \centering
    \includegraphics[width=0.99\columnwidth]{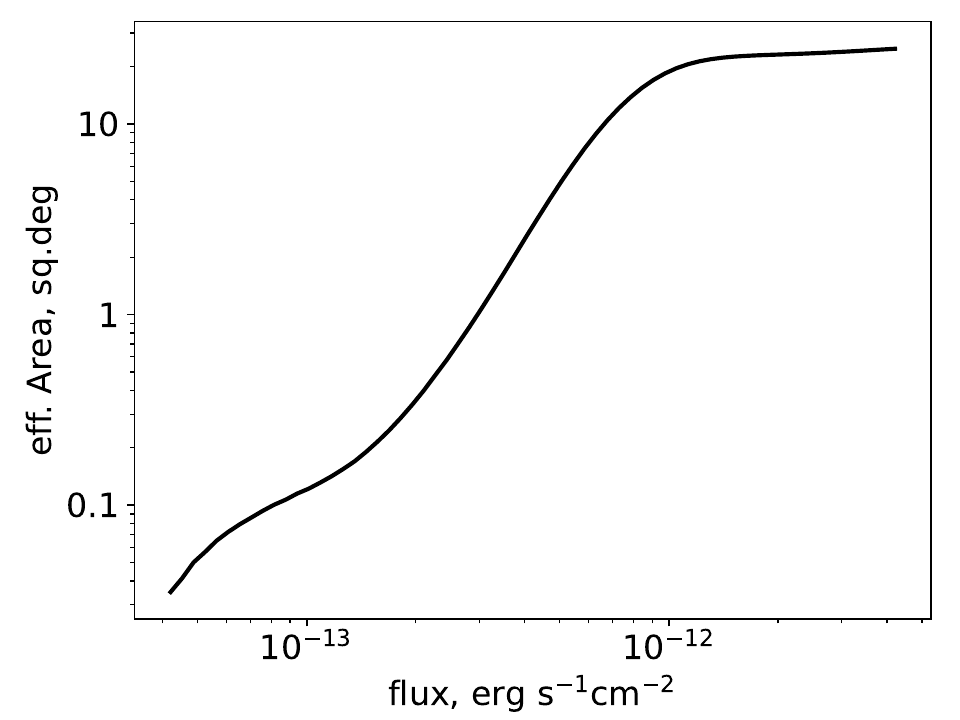}
    \caption{$A_{\rm eff} = \int \eta d\Omega$ -- , the integral of the detection completeness, ($\eta$ -- is the detection probability of an X-ray source with a given flux from a given sky direction) versus source flux over the entire survey area.}
    \label{fig:compl}
\end{figure}

The sensitivity of the survey (i.e., the X-ray flux from a source detected with a given probability) for the algorithm being used depends on the integral characteristic determined by the background conditions and the deviation of the source from the optical axis at each time of observation. At a constant background count rate observed during the survey, the sensitivity at a point is approximately proportional to the vignetted exposure. The integral of the detection completeness as a function of the X-ray flux from the source is shown in Fig.  \ref{fig:compl}. The median sensitivity of our survey at 50\% completeness is $8\times10^{-13}$~erg~s$^{-1}$~cm$^{-2}$.

\section{CATALOG OF DETECTED X-RAY SOURCES}
\label{sec:sources}

%================================================
\begin{figure*}
    \centering
        \includegraphics[width=\textwidth,trim={1.5cm 6.5cm 1.7cm 5.8cm},clip]{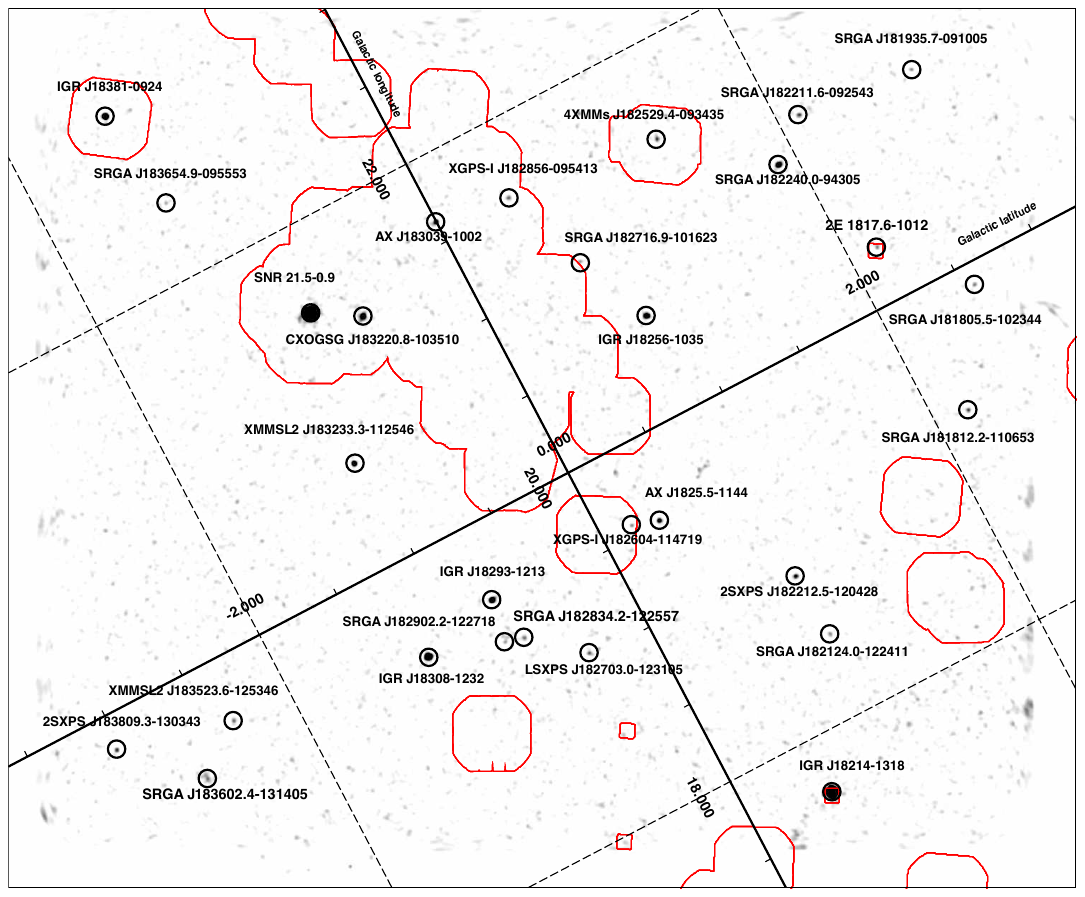}
    \caption{Detection probability map of point sources  ($\Delta \ln{L}$, search map) in the L20 field.
    All of the objects detected in the survey for which $\Delta \ln{L}>$11.4. The new sources (discovered by the ART-XC telescope) have the prefix “SGRA” in the name. The red contours mark the boundaries of the XGPS/XMM-Newton survey and the regions of other deep point XMM-Newton observations in the L20 field.
    }
    \label{fig:MAP}
\end{figure*}
%================================================

The catalog of sources detected by the ART-XC telescope in the L20 field during the observations in October 2019 includes 29 objects. Figure\,\ref{fig:MAP} presents an ART-XC image of the L20 sky region in the 4--12 keV, energy band with an indication of the detected sources. The list of all sources with coordinates, fluxes, and other information is given in Table \,\ref{tab:srclist} and its content is described below.

{ Column (1) ``Id''} — he source number in the catalog.

{ Column (2) ``Name''} -- the source name in the catalog. The prefix “SRGA” indicates that the source was detected by the SRG/ART-XC telescope, and the succeeding numerical designation corresponds to the source coordinates in the sky (J2000).

{ Columns (3,4) ``RA, Dec''} -- the equatorial coordinates of the source (J2000).

{ Column (5) ``Position error''} — the statistical error in arcseconds for the source coordinates.

{ Column (5) ``Flux''} -- the time-averaged flux from the source in the 4--12 keV energy band and the corresponding error.

{ Column (6) ``$\Delta \ln{L}$''} -- the source detection probability (for more details, see the text).

{ Column (7) ``Common name''} -- the generally known name of the source (if it has already been discovered and studied previously).

{ Column (8) ``Class''} -- the astrophysical class of the object.

{ Column (9) ``References''} -- the references to the papers in which the nature of the object was established for the first time and to the papers/catalogs where the discovery of the source is reported for the first time are given for the well-studied and poorly studied sources, respectively. 

{ The following abbreviations are used: AGN—
active galactic nuclei; HMXB -- high-mass X-ray
binaries; LMXB -- low-mass X-ray binaries; CV— cataclysmic variables; IP -- intermediate polars; SNR--supernova remnants.

To reveal the previously known X-ray sources in
the catalog of the L20 survey, we cross-matched it with the catalog of sources detected by the ART-XC telescope during the first year of the SRG all-sky survey (in December 2019 -- December 2020; Pavlinsky et al. 2022) and the source catalogs of other X-ray missions. More specifically, we used the 4XMM-DR13 catalog of the XMM-Newton observatory \citep{2020A&A...641A.136W}, the CSC2 catalog of the Chandra observatory \citep{evans10csc}, the catalog of the ASCA Galactic plane survey \citep{2001ApJS..134...77S}, and the catalogs of the INTEGRAL \citep{2022MNRAS.510.4796K,2006ApJ...636..765B} and Swift {\it Swift} \citep{2020ApJS..247...54E,LSXPS} observatories.

%=================================================================================
\newgeometry{top=110mm, bottom=3mm}
\begin{landscape}
\thispagestyle{empty}
\begin{table*}
    \caption{The catalog of sources detected in the L20 field by the SRG/ART-XC telescope}
    \label{tab:srclist}
   \footnotesize
\begin{tabular}[t]{cccccccccc} 
\hline

Id & Name & RA, & Dec, & Position error, & Flux, & $\Delta \ln{L}$ & Common name & Class & References \\
 & & deg & deg & 68\% (90\%), \arcsec &$\times 10^{-12}$, erg s$^{-1}$ cm$^{-2}$ &  &  &  &  \\

\hline

N001&{\bf SRGA\,J181805.5-102344}&274.5229&-10.3957&17.1(24.7)&0.55$^{+0.13}_{-0.10}$&14.8&&&\\
N002&{\bf SRGA\,J181812.2-110653}&274.5508&-11.1147&14.3(20.9)&0.82$^{+0.19}_{-0.11}$&19.9&&&\\
N003&{\bf SRGA\,J181935.7-091005}&274.8987&-9.1679&19.8(28.4)&0.89$^{+0.32}_{-0.23}$&12.2&&&\\
N004&SRGA\,J182022.1-101106&275.0919&-10.1851&20.1(28.9)&0.46$^{+0.18}_{-0.13}$&12.0&2E\,1817.6-1012&T Tauri& 1\\
N005&SRGA\,J182120.2-131836&275.3343&-13.3100&8.3(12.1)&33.08$^{+1.03}_{-0.93}$&1767.2&IGR\,J18214-1318&HMXB&2\\
N006&{\bf SRGA\,J182124.0-122411}&275.3501&-12.4029&19.0(27.4)&1.05$^{+0.42}_{-0.18}$&11.8&&&\\
N007&{\bf SRGA\,J182211.6-092543}&275.5482&-9.4286&18.2(25.6)&0.82$^{+0.27}_{-0.09}$&19.1&&&\\
N008&SRGA\,J182212.8-120423&275.5533&-12.0731&13.3(19.0)&1.61$^{+0.25}_{-0.22}$&32.6&2SXPS\,J182212.5-120428&&3\\
N009&{\bf SRGA\,J182240.0-094305}&275.6667&-9.7181&13.0(18.6)&1.95$^{+0.34}_{-0.27}$&63.9&&&\\
N010&SRGA\,J182525.1-114529&276.3545&-11.7582&15.3(21.4)&1.50$^{+0.22}_{-0.16}$&38.8&AX\,J1825.5-1144&&4\\
N011&SRGA\,J182529.5-093417&276.3731&-9.5714&18.6(27.3)&0.95$^{+0.18}_{-0.16}$&21.4&4XMMs\,J182529.4-093435&&5\\
N012&SRGA\,J182543.7-103501&276.4321&-10.5837&11.7(16.9)&2.13$^{+0.32}_{-0.28}$&72.7&IGR\,J18256-1035&LMXB&6\\
N013&SRGA\,J182603.5-114708&276.5146&-11.7855&16.2(23.4)&1.11$^{+0.35}_{-0.22}$&13.9&XGPS-I\,J182604-114719&&7\\
N014&SRGA\,J182704.0-123106&276.7668&-12.5185&16.3(23.7)&0.78$^{+0.19}_{-0.14}$&17.0&LSXPS\,J182703.0-123105&&8\\
N015&{\bf SRGA\,J182716.9-101623}&276.8206&-10.2731&18.1(26.8)&0.56$^{+0.25}_{-0.11}$&11.7&&&\\
N016&{\bf SRGA\,J182834.2-122557}&277.1424&-12.4324&23.3(32.3)&0.60$^{+0.27}_{-0.06}$&12.0&&&\\
N017&SRGA\,J182856.8-095429&277.2366&-9.9081&20.6(29.3)&0.75$^{+0.24}_{-0.18}$&13.9&XGPS-I\,J182856-095413&&7\\
N018&{\bf SRGA\,J182902.2-122718}&277.2590&-12.4549&17.0(24.7)&0.65$^{+0.23}_{-0.11}$&12.5&&&\\
N019&SRGA\,J182920.4-121254&277.3350&-12.2149&12.1(17.1)&2.71$^{+0.20}_{-0.21}$&89.3&IGR\,J18293-1213&CV/IP&9\\
N020&SRGA\,J183037.9-100247&277.6578&-10.0464&13.9(19.6)&1.57$^{+0.29}_{-0.19}$&43.2&AX\,J183039-1002&AGN&10\\
N021&SRGA\,J183050.3-123221&277.7096&-12.5390&10.1(14.5)&4.35$^{+0.43}_{-0.27}$&205.3&IGR\,J18308-1232&CV/IP&11\\
N022&SRGA\,J183220.1-103508&278.0836&-10.5855&14.1(19.9)&1.64$^{+0.27}_{-0.14}$&54.8&CXOGSG\,J183220.8-103510&&12\\
N023&SRGA\,J183233.7-112547&278.1403&-11.4298&12.3(17.6)&2.05$^{+0.28}_{-0.21}$&73.3&XMMSL2\,J183233.3-112546&&13\\
N024&SRGA\,J183333.6-103403&278.3901&-10.5674&7.8(10.3)&$\sim$25$^*$&2473.1&SNR\,21.5-0.9&SNR/Pulsar&14\\
N025&SRGA\,J183525.2-125404&278.8551&-12.9012&17.3(24.7)&1.01$^{+0.30}_{-0.18}$&15.4&XMMSL2\,J183523.6-125346&&13\\
N026&{\bf SRGA\,J183602.4-131405}&279.0099&-13.2348&21.5(30.9)&1.00$^{+0.28}_{-0.21}$&12.2&&&\\
N027&{\bf SRGA\,J183654.9-095553}&279.2288&-9.9315&16.8(24.7)&0.77$^{+0.30}_{-0.14}$&11.4&&&\\
N028&SRGA\,J183810.0-130336&279.5415&-13.0601&14.5(21.5)&0.81$^{+0.20}_{-0.12}$&23.7&2SXPS\,J183809.3-130343&&3\\
N029&SRGA\,J183818.7-092552&279.5781&-9.4311&10.7(15.2)&2.80$^{+0.36}_{-0.27}$&143.8&IGR\,J18381-0924&AGN&15\\
\hline
%\input{ART_L20_survey/src_tab_noidx}
%\end{longtable}
%\noalign{\vskip 4pt\hrule}
\end{tabular}

\begin{flushleft}

$^*$ The source search algorithm used in this paper was developed to search for point sources. This source is an extended one and, therefore, we admit some inaccuracy in determining its flux (for more details, see the text). 

(1) \cite{1981ApJ...243L..89F}, (2) \cite{2009ApJ...698..502B}, (3) \cite{2020ApJS..247...54E}, (4) \cite{2001ApJS..134...77S}, (5) \cite{2020A&A...641A.136W}, (6) \cite{2013A&A...556A.120M}, (7) \cite{2004MNRAS.351...31H}, (8) \cite{LSXPS}, (9) \cite{2016MNRAS.461..304C}, (10) \cite{2023A&A...671A.152M}, (11) \cite{2012A&A...542A..22B}, (12) \cite{evans10csc},(13) \cite{2018yCat.9053....0X}, 
  (14) \cite{1981ApJ...248L..23B}, (15) \cite{2016ApJ...816...38T}.

\end{flushleft}

\end{table*}
\end{landscape}
\restoregeometry 
%=================================================================================

\subsection{Known Sources in the Survey}

The L20 field under consideration was previously observed completely or partially by other X-ray observatories within the framework of survey programs and point observations. Therefore, 18 of the 29 sources detected by the ART-XC telescope were already known previously as X-ray sources, while 8 of them were previously classified using multiwavelength observations. Table\,\ref{tab:knowns} gives the statistics on these objects, and even this small sample is seen to be highly varied. For comparison, cataclysmic variables prevail in the Galactic bulge region, where the ART-XC telescope made a similar survey \citep{Semena_bulge}.

It should be emphasized that the algorithm used to detect the sources was developed to search for point sources \citep{Semena_algo}. An extended emission can also produce a signal exceeding the detection threshold, but the detection statistics and fluxes obtained in this case will be improper. Extended objects with sizes of the order of the point spread function (PSF) of the telescope \citep[$\approx 53\arcsec$, on average, for the survey mode,][]{2021A&A...650A..42P}. For such sources the detection threshold will be approximately the same as that for point sources, while the flux being estimated will be close to the integral of the source surface brightness. An example of such a detection is the supernova remnant SNR\,21.5--0.9 — one of the brightest sources in the L20 field with a size of $\approx 85\arcsec$ \citep{2010ApJ...724..572M}. 

%\newpage % return to normal page
%\advance\voffset by -5cm
%\advance\footskip by -1cm

%%%%%%%%%%%%%%%%

%================================================
\begin{table}
    \caption{Statistics of the sources of known nature in the L20 field detected in the ART-XC survey}
    \label{tab:knowns}
    \centering
        \centering
    \vskip 3mm
    \renewcommand{\arraystretch}{1.5}
    \renewcommand{\tabcolsep}{0.15cm}
%    \scriptsize
    \begin{tabular}{l|c}
    \hline
         Class & Number \\
         \hline
         AGN  & 2 \\
         HMXB & 1 \\
         LMXB & 1 \\
         CV/IP   & 2 \\
         T Tauri & 1 \\
         SNR/Pulsar  & 1 \\
         \hline
    \end{tabular}

\end{table}
%================================================

~\\

\subsection{New and Unclassified Sources}

Apart from the known X-ray sources, the ART-XC telescope has detected 11 new objects in the L20 field (boldfaced in Table \ref{tab:srclist}). Most of them are located in regions that were not covered by deep observations of other X-ray observatories. 

In particular, the most in-depth study covering part of the L20 field was previously carried by the XMM-Newton observatory within the XMM-Newton Galactic Plane Survey (XGPS) program (\citealt{2004MNRAS.351...31H}). The red contours in Fig. \ref{fig:MAP} indicate the XGPS region and the zones of other XMM-Newton point observations. All of the new sources discovered in the ART-XC survey (designated as SRGA) lie outside these contours.

We carried out a preliminary search for possible counterparts in other energy bands for all of the new X-ray sources from the L20 survey. In several cases, this allowed preliminary conclusions about their nature to be drawn. For example, by comparing the list of ART-XC objects with the catalog of the VLASS radio sky survey \citep{2021ApJS..255...30G}, we established that fairly bright radio sources fall into the error circles of four new X-ray sources from the L20 field. Moreover, all these radio sources coincide with sources from the Spitzer \citep{GLIMPSE} or 
WISE \citep{2021ApJS..253....8M}, infrared surveys, with the infrared colors of these counterparts (see Table ~\ref{tab:pAGNs}) being typical for AGNs \citep{Stern_W1W2_AGN}. 
Thus, these four sources are most likely AGNs. Note that the probable radio/infrared counterpart of the source SRGA\,J182124.0$-$122411 was already classified previously as a quasar and was entered into the Large Quasar Astrometric Catalogue 3 (LQAC-3, \citealt{2015A&A...583A..75S}).

Among the objects detected by the ART-XC telescope in the L20 field are also several previously known X-ray sources of unknown nature. According to the criteria described above, two of them also resemble AGNs (see the lower part of Table \ref{tab:pAGNs}). with the soft X-ray counterpart (XGPS-I\,J182856$-$095413) of one of them (SRGA\,J182856.8$-$095429) having already been considered previously as an extragalactic object based on an analysis of the XMM-Newton X-ray data \citep{2010A&A...523A..92M}. 

The second source -- SRGA\,J183220.1$-$103508 -- is located approximately at $20\arcmin$ from SNR\,21.5-0.9, which has been repeatedly observed previously with different X-ray telescopes as a calibration target. In the Chandra \citep{2016ApJS..224...40W} and XMM-Newton \citep{2020A&A...641A.136W} catalogs this source, known under the name CXOGSG\,J183220.8$-$103510, is marked as extended. In addition, it coincides with a bright (1.1 Jy) radio source. The set of these characteristics and the available observational data determine the heightened interest in this object and allow its nature to be studied in more detail. 

\subsection{SRGA\,J183220.1$-$103508}

To investigate the morphology of the source, we stacked the Chandra images obtained in 2000 (ObsIDs: 1842, 1843) in which the object was close to the telescope’s optical axis. In the combined photon image (Fig.~\ref{fig:clust}) it is clearly seen that the source consists of two components, point and extended ones, with the position of the point X-ray source being close to the position of the radio source whose angular
size determined by VLBI is much smaller than $1\arcsec$ \citep{2022MNRAS.515.1736K,witt23}. At the same time, the contribution of the point source to the total flux is about 50$\%$. 
%================================================
\begin{figure}
    \centering
 
        \includegraphics[width=1\columnwidth,clip]{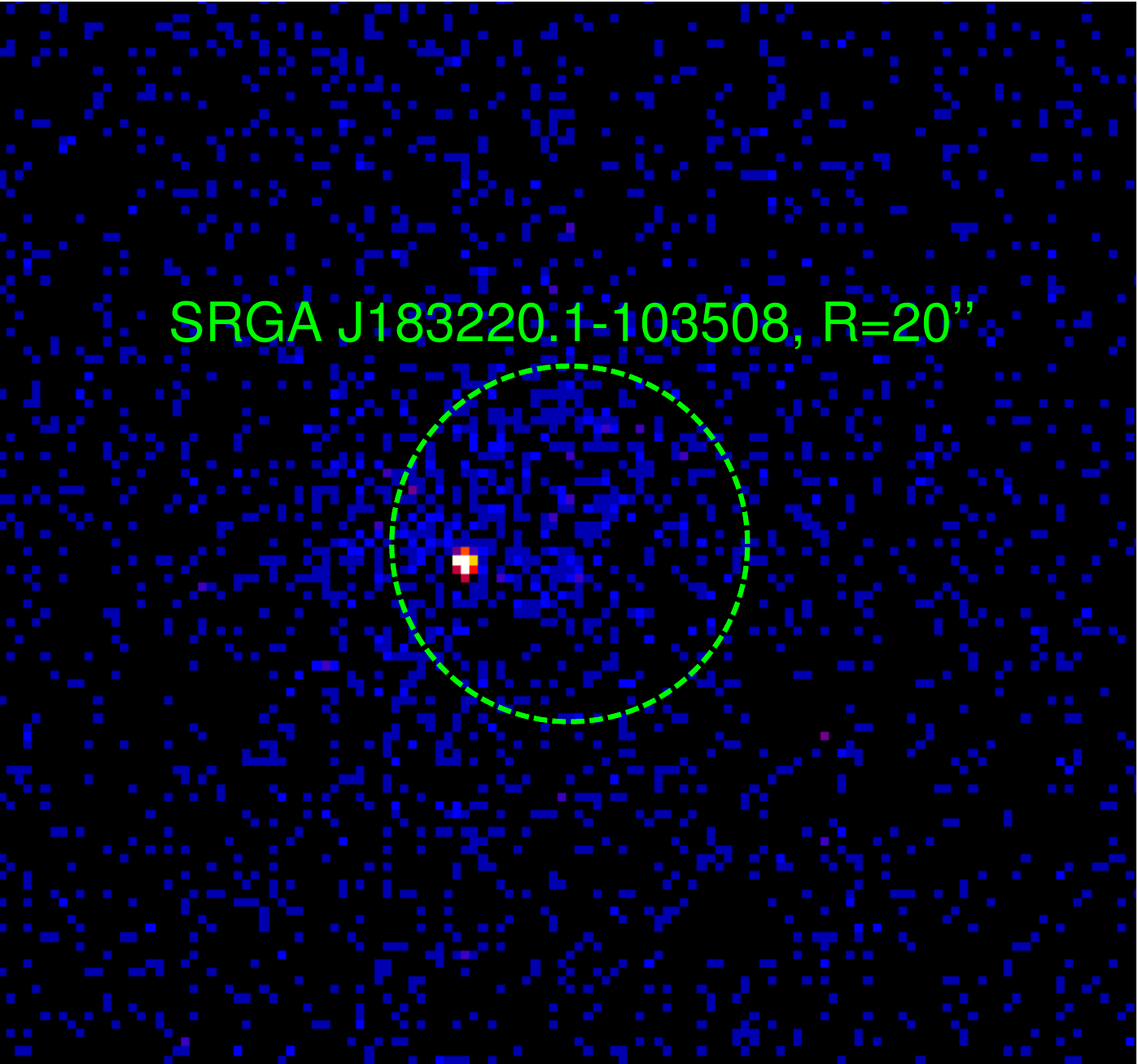}
    \caption{The field of SRGA\,J183220.1$-$103508 from Chandra data. A photon image in the 0.5--7 keV, energy band with a pixel size of $1\arcsec$. The green circle indicates the source position error circle from ART-XC data.}
    \label{fig:clust}
\end{figure}
%================================================

However, in the Chandra data the photons are not enough to obtain a high-quality spectrum of the extended component. Therefore, we invoked two additional long-term XMM-Newton observations  \xmm (ObsID: 0122700301, 0122700801) in which the source was at a distance of about $10\arcmin$ from the optical axis. Since the angular size of the extended component ($R_{c}\approx15\arcsec$, Webb et al. 2020) is comparable to the characteristic PSF width at such a distance from the axis, we extracted its spectra in circular apertures of radius $R=50\arcsec$ from the data of all three telescopes (EPIC-pn, MOS1, MOS2) by neglecting its extent. This can affect the absolute flux normalization but should not exert a significant influence on the spectral shape that is primarily of interest to us. For the subsequent analysis we added the spectra obtained in the two observations separately with the EPIC-pn camera and separately with the MOS1 and MOS2 cameras. 

Initially, we fitted both XMM-Newton spectra
simultaneously the power-law model with absorption \textsc{tbabs*(pow)} by also including the cross-
normalization constant. We used the photoionization
cross sections from \citet{vern96} and the
elemental abundances from \citet{wilms00}. This
model describes poorly the data ($\chi^{2} = 588$ per 227 degrees of freedom) and requires adding the emission
line at an energy of about 6 keV. Its addition improves
significantly the quality of the fit ($\chi^{2} = 269$ per 227 degrees of freedom), with the energy of the narrow line being $E_{\rm line} = 5.96^{+0.01}_{-0.01}$ keV and its equivalent width being about 0.6 keV. In contrast, the continuum is fairly hard, with a photon index $\Gamma = 1.8\pm0.2$, in this case, the absorption column density $N_{\mathrm{H}} = 10.8^{+0.7}_{-0.8}~\times 10^{22}$ cm$^{-2}$.  

Emission lines are often observed in supernova remnants  \citep{katsuda23snr}, but the measured energy does not correspond to the known energies of characteristic lines. Thus, the galactic nature of the source can be ruled out and, hence, it is an extragalactic object. In that case, it could be assumed that the point source is an AGN at $z\approx0.08$, while the emission line is the fluorescent Fe $K\alpha$, line whose equivalent width in Compton-thick AGNs can exceed 1 keV \citep{2010ApJ...725.2381L}.

Since the XMM-Newton angular resolution is not enough to reliably separate the contribution of the point source to the total spectrum, we separately extracted its Chandra spectrum in an aperture of radius $2.5\arcsec$. In the spectrum of the central source there is no evidence for an emission excess near 6 keV. Then, another possible explanation is that SRGA\,J183220.1$-$103508 is a galaxy cluster at redshift $z\approx0.12$. The observed emission line can then be the FeXXV line (at 6.7 keV in the source rest frame) being produced in a hot plasma. We fitted the combined XMM-Newton and Chandra spectrum by the two-component model that consists of an optically thin hot plasma spectrum describing the extended emission and a power law with reflection responsible for the contribution of the point source: \textsc{tbabs*(C$_{1}$*apec + pexrav)}. The normalization constant was taken to be $C_{1}=0.05$ for the Chandra spectrum and $C_{1}=1$ for XMM-Newton and was determined as the contribution of the extended emission to the spectrum of the point source.

%================================================
\begin{figure*}
    \centering
        \includegraphics[width=1.1\columnwidth,clip]{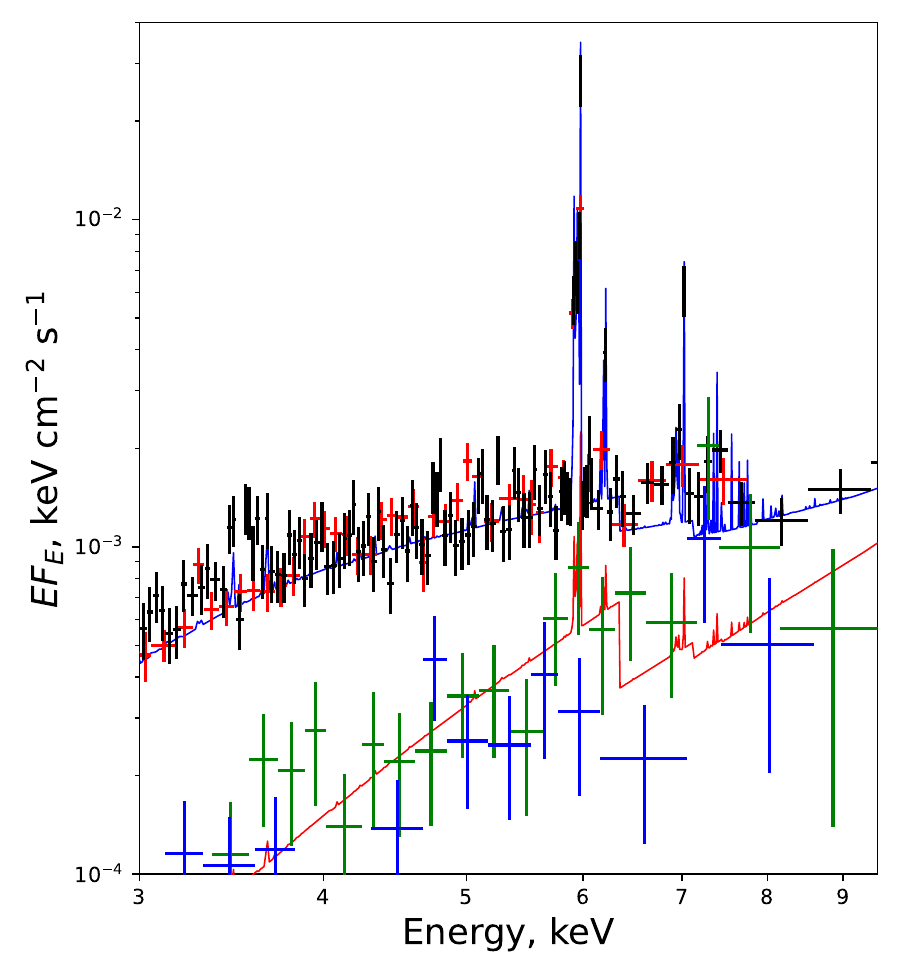}
    \caption{The spectrum of SRGA\,J183220.1$-$103508 from XMM-Newton and Chandra data: the black color indicates the EPIC-pn data, the red color indicates the MOS1 and MOS2 data, the blue and green colors indicate the Chandra spectrum of the central source obtained in observations 1842 and 1843, respectively. The solid lines indicate the best fit (for details, see the text).}
    \label{fig:xspe}
\end{figure*}
%================================================

This two-component model describes satisfactorily the data (Fig. ~\ref{fig:xspe}, $\chi^{2} = 415$ per 407 degrees of freedom); the redshift is determined as $z=0.121\pm0.001$, the plasma temperature is 
$kT=4.4^{+0.6}_{-0.5}$ keV, the heavy-element abundance relative to the solar one is $Z=0.65\pm0.10$, and the absorption column density is $N_{\mathrm{H}}=(7.3\pm0.4)\times10^{22}$ cm$^{-2}$. A shortage of data does not allow the parameters of the spectrum for the point source to be determined reliably, but it can be described by the reflected component alone (assuming the spectrum of the incident radiation with a slope $\Gamma=1.5$ and a cutoff energy $E_{\rm cut}=300$ keV). Such spectra are typical for Compton-thick AGNs (see, e.g., \citet{2019AstL...45..490S}). The total luminosity of the extended component in the standard 0.5--10 keV
X-ray energy band is $L_{X} = 2\times10^{44}$ erg s$^{-1}$, in that case, typical for galaxy clusters with such a temperature  \citep{2016A&A...592A...3G}. The small observed size of the cluster  ($R\approx 30$ kpc) may be associated with strong absorption on the line of sight in the Galaxy. Since from the available data we cannot reliably estimate the intrinsic absorption in the central source, we can only set a lower limit on its total luminosity $L_{X} \gtrsim 3 \times 10^{43}$ erg s$^{-1}$.

Thus, based on the available data, we can draw the preliminary conclusion that SRGA\,J183220.1$-$103508 is a radio-loud AGN located in a galaxy cluster at $z=0.121$. To confirm or refute this hypothesis, it is desirable to carry out additional X-ray and infrared observations to measure more accurately the parameters of the extended X-ray component (primarily its size) and to identify the cluster galaxies.

The remaining new or poorly studied objects in the catalog of ART-XC L20 field sources have no clear features that would allow their nature to be judged and require a more in-depth study that is beyond the scope of this paper. In particular, it is required to improve the positions of the new sources in the sky for which purpose it is necessary to carry out observations using X-ray telescopes with a higher angular resolution. 

%================================================
\begin{table*}[]
    \caption{Possible radio counterparts of SRGA sources from the L20 field survey from VLASS data and their infrared colors from GLIMPSE/Spitzer (or CatWISE2020/WISE, marked) data}
    \label{tab:pAGNs}
    \centering
    \vskip 2mm
    \renewcommand{\arraystretch}{1.5}
    \renewcommand{\tabcolsep}{0.12cm}
    \scriptsize
\begin{tabular}{lllllcc}
\hline

Name & radio    &  RA  & DEC & Flux$_{VLASS}$ & I1-I2 & I3-I4 \\
 {}     & counterpart & deg  & deg & mJy    &     (or W1-W2)          &        \\ 
\hline
SRGA\,J181812.2-110653 & VLASS1QLCIR\,J181811.98-110659.3 & 274.5499 & --11.1165 & 281.52\,$\pm$\,2.84 & 1.09\,$\pm$\,0.19 & 0.56\,$\pm$\,0.18 \\

SRGA\,J181935.7-091005 & VLASS1QLCIR\,J181934.70-090959.9 & 274.8946 & --9.1667 & 2.22\,$\pm$\,0.45 & 1.26\,$\pm$\,0.06$^{W}$ & - \\

SRGA\,J182124.0-122411 & VLASS1QLCIR\,J182123.27-122412.6$^{Q3}$ & 275.3470 & --12.4035 & 29.14\,$\pm$\,0.32 & 0.85\,$\pm$\,0.13 & 0.55\,$\pm$\,0.14 \\

SRGA\,J183602.4-131405 & VLASS1QLCIR\,J183600.93-131352.0 &279.0039& --13.2311&5.86\,$\pm$\,0.35 &0.74\,$\pm$\,0.25&-\\

\hline

SRGA\,J182856.8-095429 & - & 277.2337$^{X}$& --9.9038$^{X}$ & - & 0.90\,$\pm$\,0.10 & 0.71\,$\pm$\,0.10 \\

SRGA\,J183220.1-103508 & VLASS1QLCIR\,J183220.84-103510.9$^{Q3}$ & 278.0869 & --10.5864 & 1161.25\,$\pm$\,2.89 & 0.83\,$\pm$\,0.15 & 0.99\,$\pm$\,0.21\\

\hline
    \end{tabular} 

\begin{flushleft}

$^{W}$ -- from the CatWISE2020 \citep{2021ApJS..253....8M};

$^{Q3}$ -- marked as a quasar in the Large Quasar Astrometric Catalogue 3 (LQAC-3, \citealt{2015A&A...583A..75S}); 
$^{X}$ -- the coordinates from the 4XMM DR13 catalog and not from VLASS as for other objects.
\end{flushleft}

\end{table*}

\subsection{Variability of Sources}

The strategy of SRG/ART-XC observations, whereby the L20 field was covered by four scans for  $\sim$4 days, allows us to investigate the variability and flaring activity of the detected sources. Having compared the fluxes from the sources in different scans, we revealed no significant variability in any of them with the exception of IGR\,J18214$-$1318.

This source, which was previously classified as a high-mass X-ray binary \citep{2009ApJ...698..502B}, was detected twice during the ART-XC observations in a bright state with a flux of $(5-8)\times10^{-11}$ \fu, which in both cases was preceded by a significant (up to an order of magnitude) decrease in the X-ray flux (Fig.~\ref{fig:LC}). However, based on these data alone, it is impossible to conclude whether these variations are associated with the orbital periodicity of the system \citep{2020MNRAS.498.2750C} or with the strong intrinsic flaring activity \citep{2017ApJ...841...35F}.

For other L20 field sources that were previously detected by the XMM-Newton observatory we compared the fluxes in the 4--12 keV energy band obtained in this paper with the fluxes from the 4XMM (DR13) catalog in a similar energy band, 4.5--12 keV. Based on the information collected in Table  \ref{tab:L20_vs_XMMdr13}, it can be concluded that most of these sources are variable on long time scales (several years). The X-ray fluxes from some of them changed by up to three times during the ART-XC survey relative to the XMM-Newton observations.

Note also that during the L20 survey the ART-XC telescope detected no emission from the known burster  SAX\,J1828.5$-$1037 \citep{2002A&A...392..885C}. This is the only bright source (with a flux $>10^{11}$ \fu\ in the 4XMM-DR13 catalog) among all of the known ones in this sky region that was not detected in our survey. Obviously, in this period the source did not exhibit any flaring activity.

%================================================
\begin{figure}[t]
    \centering
 
        \includegraphics[width=1\columnwidth,trim={1.3cm 7.5cm 1cm 3cm},clip]{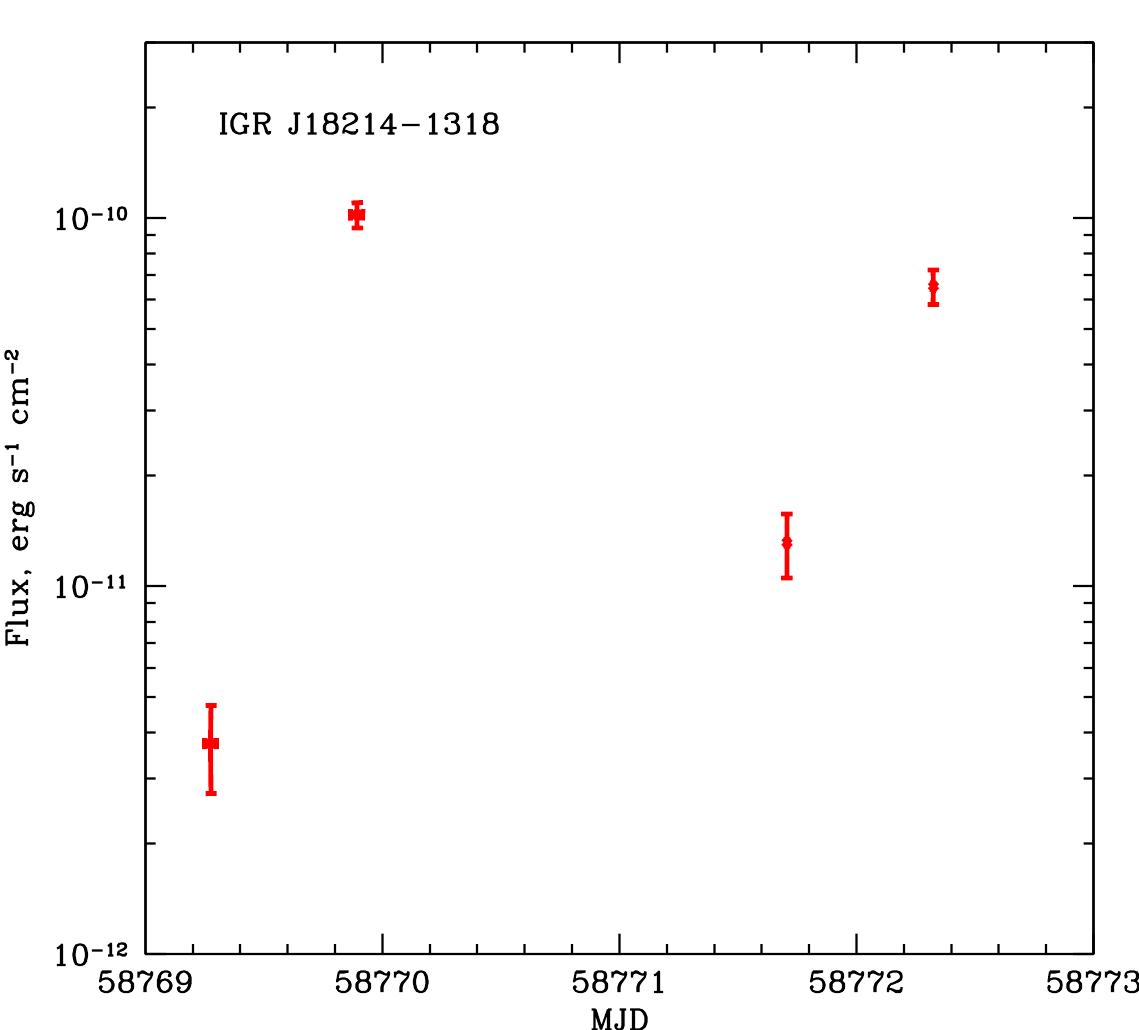}
   
    \caption{The light curve of IGR\,J18214-1318 obtained by the ART-XC telescope during the L20 field observation. Each point in the figure corresponds to the mean flux from the source in a separate “serpentine” field scan.}
    \label{fig:LC}
\end{figure}
%================================================

%================================================
\begin{table}
    \caption{Comparison of the time-averaged fluxes (in units of $10^{-12}$ erg s$^{-1}$ cm$^{-2}$) from the L20 field sources based on data from the ART-XC telescope and data from the 4XMM-DR13 catalog of the XMM-Newton observatory (in the 4--12 and 4.5--12 keV energy bands, respectively)}.
    \label{tab:L20_vs_XMMdr13}
    \centering
    \vskip 2mm
    \renewcommand{\arraystretch}{1.5}
    \renewcommand{\tabcolsep}{0.15cm}
    \scriptsize
    \begin{tabular}{l|c|c}
    \hline
         Source & Flux,& Flux,\\
          &ART-XC&4XMM \\

         \hline
    IGR\,J18214-1318  & 33.08$^{+1.03}_{-0.93}$ & 13.00\,$\pm$\,0.09\\
    AX\,J183039-1002  & 1.57$^{+0.29}_{-0.19}$ & 1.50\,$\pm$\,0.05 \\
    IGR\,J18308-1232  & 4.35$^{+0.43}_{-0.27}$ & 5.70\,$\pm$\,0.06 \\
    IGR\,J18381-0924  & 2.80$^{+0.36}_{-0.27}$ & 1.80\,$\pm$\,0.04 \\
    AX\,J1825.5-1144  & 1.50$^{+0.22}_{-0.16}$ & 1.20\,$\pm$\,0.08 \\
    XGPS-I\,J182604-114719  & 1.11$^{+0.35}_{-0.22}$ & 0.38\,$\pm$\,0.03 \\
    4XMMs\,J182529.4-093435 & 0.95$^{+0.18}_{-0.16}$ & 0.34\,$\pm$\,0.01\\
    XGPS-I\,J182856-095413 &0.75$^{+0.24}_{-0.18}$ & 2.10\,$\pm$\,0.12\\
    CXOGSG\,J183220.8-103510 & 1.64$^{+0.27}_{-0.14}$ & 5.50\,$\pm$\,0.08\\
         \hline
    \end{tabular}

\end{table}
%================================================

\section{CONCLUSIONS}
\label{sec:summary}
From the results of the L20 Galactic field survey conducted in October 2019 with the Mikhail Pavlinsky ART-XC telescope onboard the Spectrum–RG X-ray observatory we can conclude that the chosen strategy of observations demonstrated its optimality from the standpoint of performing deep homogeneous surveys of extensive sky fields. The L20 field with a total area of $\simeq 24$ sq. degree was scanned with a median sensitivity of $8\times10^{-13}$ erg s$^{-1}$ cm$^{-2}$ (at 50\% detection completeness). This allowed 29 X-ray sources to be detected in the 4--12 keV energy band, 11 of which were not detected previously by X-ray observatories. According to the available multiwavelength photometric data, four of them are probably extragalactic in nature and belong to the class of active galactic nuclei.

We also studied in detail the properties of one of the previously known sources detected in the survey, -- SRGA\,J183220.1$-$103508 (CXOGSG\,J183220.8$-$103510). According to our estimates, it is most likely a galaxy cluster containing a bright radio galaxy at redshift $z\simeq0.121$.

%\newpage % larger page1
%\enlargethispage{1.5cm} % more room for text or floats
%\advance\voffset by 5.5cm % reduce top margin
%\advance\footskip by 1cm % lower page number

\section{Note added in the arXiv version of the paper}

After publication of the paper, it was brought to our attention that SRGA\,J183220.1$-$103508 is likely associated with the source XMMU\,J183225.4$-$103645, whose nature had been previously investigated by \citet{2001A&A...374...66N}. Our conclusions agree with the finding of these authors that the source is a galaxy cluster at redshift $z\sim0.12$. However, we also unveiled the presence of a bright AGN within the cluster. We initially missed this association because of the offset of ${\simeq}2$\arcmin\ between SRGA\,J183220.1$-$103508 and XMMU\,J183225.4$-$103645, which significantly exceeds the 90\% localization error of 19.9\arcsec\ used in our work. Apparently, there is a significant systematic error in the coordinates of XMMU\,J183225.4$-$103645 reported by \citet{2001A&A...374...66N}, based on performance verification observations of the XMM-Newton observatory. The position of SRGA\,J183220.1$-$103508 is consistent with the latest 4XMM DR13 and Chandra CSC 2.0 catalogs data.

\section{ACKNOWLEDGMENTS}

In this study we used data from the Mikhail Pavlinsky ART-XC telescope onboard the Spectrum--Roentgen--Gamma (SRG) observatory. The SRG observatory was designed by the Lavochkin Association (enters into the Roskosmos State Corporation) with the participation of the Deutsches Zentrum fur Luft- und Raumfahrt (DLR) within the framework of the Russian Federal Space Program on the order of the Russian Academy of Sciences. The ART-XC team thanks the Roskosmos State Corporation, the Russian Academy of Sciences, and the Rosatom State Corporation for supporting the design and production of the ART-XC telescope and the Lavochkin Association and partners for the production and work with the spacecraft and the Navigator platform. 

The authors also thank Victor Muteti for valuable comments on our results.

\section{FUNDING}

This study was supported by RSF grant no. 19-12-00396 with regard to the search for and investigation of active galactic nuclei.

\section{CONFLICT OF INTEREST}

The authors of this work declare that they have no
conflicts of interest.

%\newpage

\bibliographystyle{astl}
\bibliography{biblio} 

%================================================

{\it Translated by V. Astakhov}

{\it Latex style was created by R. Burenin}

 % Don't change these lines
%\bsp    % typesetting comment
\label{lastpage}

\end{document}